\documentstyle[epsfig]{aipproc}
\newcommand{\md}{\mbox{d}}
\newcommand{\sn}{\mbox{\rm sn}}

\begin{document}
\title{Test particle acceleration\\
       by rotating jet magnetospheres}

\author{Frank M. Rieger and Karl Mannheim}
\address{Universit\"ats-Sternwarte G\"ottingen\\ 
Geismarlandstr. 11, 37083 G\"ottingen, Germany}

\maketitle
\begin{abstract}
Centrifugal acceleration of charged test particles at the base
of a rotating jet magnetosphere is considered. 
Based on an analysis of forces we derive the equation for the radial 
accelerated motion and present an analytical solution. It is shown that for 
particles moving outwards along rotating magnetic field lines, the energy 
gain is in particular limited by the breakdown of the bead-on-the-wire 
approximation which occurs in the vicinity of the light cylinder 
$r_{\rm L}$. The corresponding upper limit for the maximum Lorentz factor 
$\gamma_{\rm max}$ for electrons scales $\propto B^{2/3} \,r_{\rm L}^{2/3}$, 
with $B$ the magnetic field strength at $r_{\rm L}$, and is at most of the 
order of a $10^2-10^3$ for the conditions regarded to be typical for BL~Lac 
objects. Such values suggest that this mechanism may provide pre-accelerated 
seed particles which are required for efficient Fermi-type particle 
acceleration at larger scales in radio jets.
\end{abstract}

\section*{Introduction}
\begin{figure}[htb]
       \vspace{0cm}
       \begin{center}
         \epsfxsize=13.5cm         
          \mbox{\epsffile{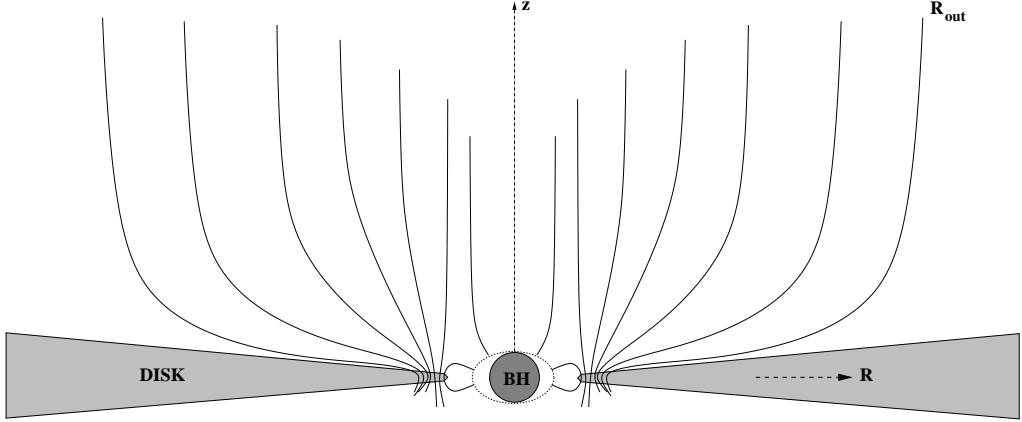}}
       \end{center}
       \vspace{0cm} 
       \caption{Simplified model topology for the asymptotic jet structure
        around a rotating black hole as expected in MHD scenarios.} 
       \label{jet}
   \end{figure}

    Rotating magnetospheres are widely believed to be responsible 
    for the relativistic jet phenomenon in active galactic nuclei (AGN) 
    \cite{blandford82,begelman94}.
    Here we adress the question whether centrifugal acceleration of
    charged test particles at the base of such a jet magnetosphere may 
    possibly produce a seed population of relativistic electrons which is 
    required for efficient particle acceleration. For, in order to explain 
    the origin of the nonthermal emission extending up to TeV energies in 
    some blazars, several acceleration processes have been proposed among 
    which Fermi-type particle acceleration mechanisms (i.e. diffusive shock 
    acceleration \cite{drury83}) are quite promising. However such kind of 
    mechanisms require a pre-accelerated seed population of electrons with 
    Lorentz factors of the order of $100$ \cite{melrose94,lesch97}.     
    It seems therefore quite interesting whether in the case of AGN
    centrifugal acceleration by rotating jet magnetosphere may potentially 
    fill this gap by providing pre-accelerated seed particles.
    For an analytical treatment, we consider the following simplified model: 
    motivated by MHD-scenarios for the origin of jets via rotating jet 
    magnetospheres \cite{blandford82,camenzind96,fendt97} (see Fig.~\ref{jet})
    a projected two-dimensional model topology is applied where the magnetic 
    field is supposed to rotate rigidly with a fraction of the rotational 
    velocity of the black hole \cite{fendt97}.  
    Test particles with rest mass $m_e$ and charge $e$ are assumed to be 
    injected at time $t_0$ and position $r_0$ with velocity $v_0$ parallel 
    to the rotating field line.
\section*{Modelling}
 Consider the forces acting on a particle in a rotating frame of reference 
 \cite{ganga96,ganga97}:
 Particles, which are injected at ($t_0$,$r_0$) with velocity $v_0$ along the 
 magnetic field line $B_r(t_0)$ experience a centrifugal force in the radial
 direction given by
 \begin{equation}
 \vec F_{\rm cf}=m_e\,\gamma\,(\vec \Omega \times \vec r)\times \vec \Omega\,,
 \end{equation} where $\gamma$ denotes the Lorentz factor and $\vec \Omega= 
 \Omega \, \vec e_z$ the angular velocity of the field. 
 Additionally, there is also a relativistic Coriolis term in the noninertial 
 frame governed by the equation 
 \begin{equation}
 \vec F_{\rm cor}=m_e\,\left(2\,\gamma \frac{\md r}{\md t} + r\,
 \frac{\md\gamma}{\md t}\right)\,(\vec e_r \times \vec \Omega)\,,
 \end{equation} which acts as a deviation-force in the azimuthal direction. In 
 the inertial rest frame the particle sees the field line bending off from its 
 initial injection position, therefore it experiences a Lorentz force ($c=1$)
 \begin{equation}
 \vec F_{\rm L}= e\,(\vec v_{\rm rel} \times \vec B)\,,
 \end{equation} where $v_{\rm rel}$ is the relative velocity between the 
 particle and the magnetic field line. Due to the Lorentz force a particle 
 tries to gyrate around the field line. Initially, the direction of the 
 Lorentz force is perpendicular to the direction of the Coriolis force, but as
 a particle gyrates, it changes the direction and eventually becomes 
 antiparallel to the Coriolis force. Hence, the bead-on-the-wire 
 approximation is valid if the Lorentz force is not balanced by the Coriolis 
 force \cite{ganga97,rieger00}. 
 In this case, the accelerated motion of the particle's guiding center 
 due to the centrifugal force may be written as
 \begin{equation}\label{radial}
 \gamma \frac{\md^2 r}{\md t^2} +\frac{\md r}{\md t} \,
   \frac{\md \gamma}{\md t} = \gamma\, \Omega^2 \,r\,,
 \end{equation} where $\gamma=(1-\Omega^2 r^2-\dot r^2)^{-0.5}$.
 The constrained motion is then given by the azimuthal components of forces 
 \begin{equation}\label{constraint}
 \frac{\md \gamma}{\md t} \leq \frac{1}{r}
    \left(\frac{B\,e\,v_{\rm{rel}}}{\,m_e\,\Omega} -2\,\gamma 
    \frac{\md r}{\md t}\right)\,.
 \end{equation}
 Generally, the bead-on-the-wire approximation is supposed to break down if 
 $F_{\rm cor}$ exceeds $F_{\rm L}$ (i.e. when $\leq$ in Eq.~\ref{constraint} 
 becomes $>$).

\section*{Results}
 Using the argument that the Hamiltonian for a bead on a relativistically 
 moving wire $H =\gamma\,m_e\,(1 - \Omega^2\,r^2)$ is a constant of motion,
 the equation for the radial accelerated motion could be reduced to a simple 
 form which has been solved analytically yielding \cite{machabeli94,rieger00} 
 \begin{equation} 
 r(t)=\frac{1}{\Omega}\, \rm{cn}(\lambda_0 - \Omega\,t)\,,
 \end{equation} where $\rm{cn}$, ($\sn$) is the Jacobian elliptic cosine 
 (sine, respectively), and $\lambda_0$ is an elliptic integral of the 
 first kind, i.e. 
 \begin{equation}
 \lambda_0=\int_0^{\phi_0}\,\frac{\md \theta}
                   {(1-\tilde{m}\,\sin^2\theta)^{1/2}}\,.
 \end{equation} with $\tilde{m}=(1-\Omega^2 r_0^2-v_0^2)/(1-\Omega^2 r_0^2)^2$.
 The Lorentz factor may then be written as
 \begin{equation}
 \gamma(t)=\frac{1}{\sqrt{\tilde m}\,[\sn(\lambda_0 - \Omega\,t)]^{2}}\,,
 \end{equation} or, if expressed as a function of the radial co-ordinate, as
 \begin{equation}\label{gamma_r}
 \gamma(r(t))=\frac{1}{\sqrt{\tilde m}\,(1-\Omega^2\,\, r(t)^2)}\,.
 \end{equation}
 Apart from radiation losses (e.g. inverse-Compton losses in the radiation
 field of the accretion disk, see \cite{rieger00}), the maximum attainable 
 Lorentz factor $\gamma_{\rm max}$ is in particular limited by the breakdown 
 of the bead-on-the-wire approximation (i.e. when the particle leaves the 
 field line and thus, acceleration becomes ineffective) in the vicinity of the
 light cylinder $r_{\rm L}$. 
 Using the definition of the Hamiltonian $H$ and Eq.~\ref{gamma_r} and 
 setting $v_{\rm rel}=c$, one may derive an upper limit 
 for the maximum Lorentz factor $\gamma_{\rm max}$ from Eq.~\ref{constraint} 
 \begin{equation}\label{gmax}
  \gamma_{\rm max} \simeq \frac{1}{\tilde{m}^{1/6}}\,
  \left(\frac{B(r_{\rm L}\,)\,e}{2\,m_e\,c^2}\, \,r_{\rm L}\,\right)^{2/3}\,,
 \end{equation} where $B(r_{\rm L})$ denotes the magnetic field strength at 
 the light cylinder and where for clarification $c$ has now been inserted. 
 For typical BL Lac conditions, i.e. a light cylinder radius 
 $r_{\rm L}\sim 10^{13}$\,m, and a field strength $B(r_{\rm L}) \sim (30-100)
 \times 10^{-4}$ T, Eq.~\ref{gmax} results in an upper limit on the maximum 
 Lorentz factor $\gamma_{\rm max} \sim (1-2.5)\times 10^3$.
\section*{Conclusions}
 The results derived in the simple toy-model presented here support flares on 
 accretion disks as providing a seed population of relativistic electrons 
 with Lorentz factors up to $\sim 10^3$ in BL Lac type objects. 
 Such pre-accelerated particles are required for models involving diffusive 
 shock acceleration of $e^+/e^-$ in relativistic jets, cf. \cite{melrose94}, 
 \cite{lesch97}. Particle acceleration by rotating jet magnetospheres
 may thus possibly represent an interesting explanation for the required
 pre-acceleration.

{\small Acknowledgement: K.M. acknowledges financial support from a 
Heisenberg-Fellowship and F.M.R. from DFG Ma 1545/2-2.}


\begin{references}
\bibitem{begelman94} Begelman, M.C., ``Magnetic Propulsion of Jets in AGN,'' 
     in {\it The Nature of Compact Objects in Active Galactic Nuclei}, 
     edited by A. Robinson, and R. Terlevich, Univ. Press, Cambridge, 1994, 
     pp. 361-367. 
\bibitem{blandford82} Blandford, R.D., and Payne, D.G., {\it MNRAS} {\bf 199}, 
     883 (1982).
\bibitem{camenzind96} Camenzind, M., ``Stationary Relativistic MHD Flows,'' 
     in {\it Solar and Astrophysical Magnetohydrodynamic Flows}, edited by
     K.C. Tsinganos, Kluwer Academic Publ., Dordrecht, 1996, pp. 699-725. 
\bibitem{drury83} Drury, L.O'C., {\it Rep. Progr. Phys.} {\bf 46}, 973 (1983).
\bibitem{fendt97} Fendt, C., {\it A\&A} {\bf 319}, 1025 (1997).
\bibitem{ganga96} Gangadhara, R.T., {\it A\&A} {\bf 314}, 853 (1996).
\bibitem{ganga97} Gangadhara, R.T., and Lesch, H., {\it A\&A} {\bf 323}, 
     L45 (1997).
\bibitem{lesch97} Lesch, H., and Birk, G.T., {\it{A\&A}} {\bf 324}, 461 (1997).
\bibitem{machabeli94} Machabeli, G.Z., and Rogava, A.D., {\it Phys. Rev. A} 
    {\bf 50}, 98 (1994).
\bibitem{melrose94} Melrose, D.B., in: Kirk, J.G., Melrose, D.B., and 
       Priest, E.R., {\it Plasma Astrophysics}, edited by A.O. Benz, and 
       T.J.-L. Couvoisier, Springer, Berlin, 1994, pp. 113-223.
\bibitem{rieger00} Rieger, F.M., and Mannheim, K., {\it A\&A} {\bf 353}, 473 
      (2000). 
\end{references}
\end{document}